\documentclass[12pt]{article}

\usepackage[pdftex]{graphicx}
\usepackage{cite}

\usepackage{multirow}%
\usepackage{amsmath,amssymb,amsfonts}%
\usepackage{amsthm}%
\usepackage{mathrsfs}%
\usepackage{xcolor}%
\usepackage{authblk}

\newcommand{\bq}{\begin{equation}}
\newcommand{\eq}{\end{equation}}
\newcommand{\bqa}{\begin{eqnarray}}
\newcommand{\eqa}{\end{eqnarray}}

\newcommand{\sss}[1]{\scriptscriptstyle{#1}}

\def\mw {M_{\sss{W}}}

\definecolor{darkblue}{rgb}{0.0, 0.0, 0.55}
\definecolor{darkgreen}{rgb}{0.0, 0.2, 0.13}
\definecolor{darkcandyapplered}{rgb}{0.64, 0.0, 0.0}
\definecolor{color1}{rgb}{0.7,0,0}
\definecolor{color2}{rgb}{0,0,0.7}
\definecolor{color3}{rgb}{0,0,1}
\definecolor{color4}{rgb}{0.8,0,0}
\definecolor{color5}{rgb}{0,0.7,0}

\newcommand{\MeV}{{\text{ MeV}}}
\newcommand{\GeV}{{\text{ GeV}}}

\usepackage{accents}

\def\mw  {M_{\sss{W}}}
\def\mz  {M_{\sss{Z}}}

\def\sanc{\tt {SANC}}
\def\mcsancee{\tt {MCSANCee}}
\def\renesance{\tt {ReneSANCe}}
\def\calchep{\tt {CalcHEP}}
\def\whizard{\tt {WHIZARD}}

\def\cvz{\cos{\vartheta_{Z}}}
\def\sqs{\sqrt{s}}
\def\sba{$\sigma^{\text{Born}}_{\alpha(0)}$}
\def\sbg{$\sigma^{\text{Born}}_{G_\mu}$}

\def\dbg{$\delta^{\text{Born}}_{G_\mu / \alpha(0)}$}

\def\swa{$\sigma^{\text{weak}}_{\alpha(0)}$}
\def\swg{$\sigma^{\text{weak}}_{G_\mu}$}

\def\dwg{$\delta^{\text{weak}}_{G_\mu / \alpha(0)}$}

\def\alr{A_{\rm LR}}

\begin{document}

\title{One-loop electroweak radiative corrections to polarized $e^+e^- \to \gamma Z$ process}

\author[1,2]{S.~Bondarenko}
\author[3,4]{Ya.~Dydyshka}
\author[3]{L.~Kalinovskaya}
\author[3]{L.~Rumyantsev}
\author[3]{R.~Sadykov}
\author[3,4]{V.~Yermolchyk}

\affil[1]{\small Bogoliubov Laboratory of Theoretical Physics, JINR, 
                 141980 Dubna, Moscow region, Russia}
\affil[2]{\small Dubna State University,  
                 Universitetskaya str. 19,  141980 Dubna, Russia}
\affil[3]{\small Dzhelepov Laboratory of Nuclear Problems, JINR,  
                 141980 Dubna, Moscow region, Russia}
\affil[4]{\small Institute for Nuclear Problems,
  Belarusian State University, Minsk, 220006, Belarus}

\date{}

\maketitle

\abstract{
This paper gives high-precision theoretical predictions for cross sections
of the process $e^+e^- \to  \gamma Z$ for future electron-positron colliders.
The calculations are performed using the {\sanc} system.
They include complete one-loop electroweak radiative corrections,
as well as longitudinal polarization of the initial state.
Numerical results are given for the center-of-mass system energy
range $\sqrt{s} = 250 - 1000$~GeV with various polarization degrees in the $\alpha(0)$ and $G_\mu$ EW schemes.
This study is a contribution to the research program of the 
CEPC project being under development in China.
}

\section{Introduction}

Clean experimental conditions, negligible pile-up and
controlled centre-of-mass energy, 
part-per-mil accuracy for cross sections for signal and background processes, and well-understood backgrounds for final states are assumed at future $e^+e^-$ colliders.
All this allows us to search for new physics by accurately measuring deviations from the Standard Model
\cite{Blondel:2018aan,Blondel:2019qlh}.
However, there are  challenges that arise from the very richness of the $e^+e^-$ program.
 One needs to match theoretical accuracy with the statistical one  taking into account electroweak (EW) radiative corrections (RCs)~\cite{LEP}. 
 
Modern evaluation tools to estimate theoretical uncertainties for future
$e^+ e^-$ colliders, i.e., FCC{ee} \cite{Blondel:2018mad,FCC:2018byv},
ILC~\cite{ILC:2013jhg,Fujii:2015jha,ILC:2019gyn},
CLIC~\cite{CLICdp:2018esa},
CEPC~\cite{CEPCStudyGroup:2018ghi,Duan:2023lyp}
should be applied. 
An important advantage of linear collider ILC and CLIC is 
planned
high degree of polarization
 of electron beams. 
Polarized beams can improve opportunities for investigation of the fundamental particle properties  ~\cite{Blondel:2019qlh,MoortgatPick:2005cw,Fujii:2015jha,Blondel:2018mad,Bambade:2019fyw}.

In the  physical programs for the future $e^+e^-$ colliders 
significant attention devoted to the  
$e^+e^- \to \gamma Z$
process. 
All of them have options  to operate at centre-of-mass  energies in the range $240 - 250$ GeV,  and 
are expected to accumulate about $10^5 - 10^6$ Higgs boson events. 

At this energy the dominant Higgs production channel
is the Higgsstrahlung process $e^+e^- \to HZ$~\cite{Bondarenko:2022xmc}.
Because of clear initial $e^+e^-$ state it is possible to identify Higgs boson events independently of the decay mode, and the largest background comes from two
 processes  $e^+e^- \to \gamma Z$  and $e^+e^- \to ZZ$.

Also it is supposed 
to precisely calibrate the energy scales of the particles
using the $e^+e^- \to \gamma Z$ process
for measuring  Higgs boson coupling constants~\cite{Mizuno:2021dxw},
and for evaluating the left-right asymmetry
at the center-of-mass energy of 250 GeV
\cite{Aihara:1995iq,Mizuno:2022xuk}. 
 
An accurate measurement of the number of neutrinos
is proposed by using the radiative return to the Z boson, 
$e^+e^- \to \gamma Z$~\cite{FCC:2018byv}.
 
Here, we focus on the evaluation and analysis of the effects associated with the longitudinal polarization of the initial beams of the process $e^+e^- \to \gamma Z$.

This is the second part of the study of theoretical
uncertainties 
 of the process~\cite{Bardin:2007dz} 
\bqa
e^{+}(p_1,\chi_1) + e^{-}(p_2,\chi_2) \rightarrow
\hskip 20mm
\nonumber\\
\rightarrow  \gamma(p_3,\chi_3) + Z(p_4,\chi_4) \  (+ \gamma(p_5,\chi_5))
\label{Reac_eegZ}
\eqa
at the complete one-loop electroweak level in {\tt SANC},
with $p_i, \chi_i$ being the momentum and helicity of the $i$th particle.

In our previous paper \cite{Bardin:2007dz}
the relevant contributions for one-loop corrections,
i.e., Born, virtual and real
photon bremsstrahlung, 
to the cross section were calculated analytically,
the independence of the form factors of gauge parameters
was tested, and the stability of the result on the variation of the soft-hard
separation parameter ${\bar\omega}$ was checked.
We keep all masses and operate in full phase space.
All calculations were carried out via the {\mcsancee}~\cite{MCSANCee:2021} integrator
and the {\renesance} generator~\cite{SADYKOV2020107445}.

One-loop QED and EW RCs to the unpolarized $e^+e^- \to \gamma Z$ process
were previously calculated
in~\cite{CapdequiPeyranere:1984sj,Berends:1986yy,Bohm:1986mz,Gounaris:2002za}.
However, it is difficult to draw a direct comparison between our results
and these papers due to incomplete setups.

The article is organized as follows. 
Section 2 describes the general notations.
Numerical results and comparison are presented in Section~\ref{NumResultsComp}. 
A summary is given in Section~\ref{sect_Concl}.

\section{Differential cross section} \label{PAP}

To study the case of the longitudinal polarization with degrees $P_{e^+}$ and $P_{e^-}$,
we produce helicity amplitudes and make a formal application of Eq.~(1.15) from~\cite{MoortgatPick:2005cw}:
\begin{equation}
\sigma{(P_{e^+},P_{e^-})} = \frac{1}{4}\sum_{\chi_1,\chi_2}(1+\chi_1P_{e^+})(1+\chi_2P_{e^-})\sigma_{\chi_1\chi_2},
\label{eq1}
\end{equation}
where $\chi_{i} = -1(+1)$ corresponds to the particle $i$ with the left (right) helicity.

The cross section of the process at the one-loop level can be divided into four parts:
\begin{eqnarray}
{
  \sigma^{\text{one-loop}} =
  \sigma^{\mathrm{Born}} + \sigma^{\mathrm{virt}}(\lambda)
  + \sigma^{\mathrm{soft}}(\lambda,{\bar\omega})}{
  + \sigma^{\mathrm{hard}}({\bar\omega})},
\nonumber
\end{eqnarray}
where {$\sigma^{\mathrm{Born}}$} is the Born cross section,
{$\sigma^{\mathrm{virt}}$} is the contribution of virtual (loop) corrections, 
{$\sigma^{\mathrm{soft}}$} is the contribution of soft photon emission,
{$\sigma^{\mathrm{hard}}$} is the contribution of hard photon emission
(with energy { $E_{\gamma} > {\bar\omega}$}).
Auxiliary parameters {$\lambda$} ("photon mass") 
and {${\bar\omega}$} (soft-hard separator) 
are canceled after summation.

\section{Numerical results and comparisons} \label{NumResultsComp}

To estimate the theoretical uncertainties we consider results calculated in the $\alpha(0)$
and $G_{\mu}$ EW schemes.
Numerical results for NLO (next-to-leading-order) EW RCs involve 
the one-loop cross sections, corresponding angular and energy distributions,
the polarization effect for  the following 
longitudinally polarized states
of the positron ($P_{e^+})$ and electron ($P_{e^-}$) beams
\bqa
(P_{e^+},P_{e^-}) = (0,0),(-1,+1),(+1,-1),
\hskip 15mm
\nonumber\\
(0.3,-0.8),(-0.3,0.8),(0,-0.8),(0,0.8)
\nonumber
\eqa
and set of center-of-mass (c.m.) energies
\bqa
\sqrt{s}= 250, 500, 1000 \GeV.
\label{set_energy}
\eqa

The following setup of input parameters is used:
\begin{eqnarray}
\begin{array}{ll}
\alpha(0)= 1/137.03599976, & G_{\mu} = 1.16637 \times 10^{-5} \GeV^{-2}, \\
M_Z = 91.1876 \GeV, & \Gamma_Z  = 2.49977 \GeV,\\
M_W = 80.451495 \GeV & M_H = 125.0 \GeV, \\
m_e = 0.5109990 \MeV, & m_\mu  = 0.105658 \GeV, \\
m_\tau  = 1.77705 \GeV, &	\\
m_u = 0.062 \GeV, & m_d = 0.083 \GeV, \\
m_c  = 1.5 \GeV, & m_s  = 0.215 \GeV, 	\\
m_t = 173.8 \GeV, & m_b  = 4.7 \GeV.
\end{array}
\nonumber
\end{eqnarray}

 In the calculations, the following cuts were imposed:
 \begin{itemize}
 \item
 c.m. angular cuts for the Born, soft and virtual contributions where there 
 is only one photon in the final state $\cos{\vartheta_\gamma} \in [-0.9,0.9]$
 \item for the hard
 contribution of both photons must have a c.m. energy greater than ${\bar \omega}$;
 \item
 for the hard event to be accepted, $\cvz$ and
 at least one photon $\cos{\vartheta_{\gamma_1}}$,  $\cos{\vartheta_{\gamma_2}}$ must lie within
 the interval $[-0.9, 0.9]$.
 \end{itemize}
 It should be noted that this coincides with the criteria which were used in~\cite{Bardin:2007dz}.

\subsection{Triple comparison of the tree level results: Born and hard photon
bremsstrahlung cross sections}

\begin{table}[!ht]
\caption{
Tuned triple comparison between {\sanc} (first line), {\calchep} (second line) and {\whizard} (third line)
results for the hard bremsstrahlung contributions to polarized $e^+e^- \to \gamma Z(\gamma)$ scattering
for various degrees of polarization and energies.}
\centering
\begin{tabular}{|l|c|c|c|c|}
\hline
$P_{e^+}$, $P_{e^-}$ & $-1, -1$ & $-1, +1$ & $+1, -1$ &  $+1, +1$\\
\hline
\multicolumn{5}{|c|}{$\sigma^{\rm hard}_{e^+e^-}$, pb, $\sqrt{s} = 250$ GeV}\\
\hline
S &  2.51(1) & 69.74(1) & 110.09(1) & 2.53(1)      \\
C &  2.53(1) & 69.75(1) & 110.09(1) & 2.53(1)      \\
W &  2.53(1) & 69.75(1) & 110.07(2) & 2.53(1)      \\
\hline
\multicolumn{5}{|c|}{$\sigma^{\rm hard}_{e^+e^-}$, pb, $\sqrt{s} = 500$ GeV}\\
\hline
S &  0.74(1) & 17.04(1)  & 26.89(1)  & 0.75(1)   \\
C &  0.76(1) & 17.03(1)  & 26.88(1)  & 0.76(1)   \\
W &  0.76(1) & 17.05(1)  & 26.90(1)  & 0.76(1)   \\
\hline
\multicolumn{5}{|c|}{$\sigma^{\rm hard}_{e^+e^-}$, pb, $\sqrt{s} = 1000$ GeV}\\
\hline
S & 0.202(1)  & 4.604(1)  & 7.266(1) & 0.206(1) \\
C & 0.206(1)  & 4.603(1)  & 7.267(1) & 0.206(1) \\
W & 0.206(1)  & 4.603(1)  & 7.265(1) & 0.206(1)  \\
\hline
\end{tabular}
\label{Table:tuned_hard_SCW}
\end{table}

Here we compare for fully polarized
Born and hard photon bremsstrahlung cross sections with
the ones obtained via the
{\calchep}~\cite{Belyaev:2012qa} and {\whizard}~\cite{Kilian:2007gr,Kilian:2018onl}
codes. 
The results are given within the $\alpha(0)$ EW scheme with any photon energy $E_{\gamma} > {\bar \omega}$, 
${\bar \omega} = 10^{-4} \sqrt{s}/2$ and fixed 100\% polarized initial beams in the full phase space.

The agreement for the Born cross section was found to be excellent (we omit the corresponding table). 
Table~1 shows very good agreement between ~{\sanc} results (first row)
for the hard photon bremsstrahlung cross section contributions, 
{\calchep} results (second row) and ~{\whizard} results (third row).

\subsection{Polarized total cross sections}

\subsubsection{Energy dependence}

The corresponding unpolarized/polarized results for the Born and complete one-loop EW cross sections
as well as for relative corrections
are presented in Table~2.
Relative corrections $\delta^i$ are computed as the ratios
of the corresponding
one-loop contributions to the Born level cross section for three energies.  
We only show the components $\sigma_{-+}$ and $\sigma_{+-}$ because even in cases of a partly polarized initial state, the polarized cross sections are mainly determined by these components.


\begin{table}[!ht]
\caption{
Born and complete one-loop cross sections $\sigma$ in pb,
and relative corrections $\delta$ in \%
for the c.m. energies $\sqrt{s}=250,500,1000$~GeV,
at various  degree of polarization
of the initial particles in  the $\alpha(0)$ EW scheme.}
\centering  
\begin{tabular}{|l|c|c|c|c|c|c|c|}
\hline
$P_{e^+}$, $P_{e^-}$ & $0,0$       & $-1,+1$    & $+1,-1$ & $0.3,-0.8$ & $-0.3,0.8$ & $0,-0.8$ & $0,0.8$
\\
\hline
\multicolumn{8}{|c|}{$\sqrt{s} = 250$~GeV}\\
\hline
$\sigma^{\text{Born}}$, pb
                     & 4.094(1)  & 6.353(1) & 10.025(1) & 6.087(1) & 4.067(1) & 4.829(1) & 3.360(1)
\\
{$\sigma^{\text{NLO}}$, pb}
                     & 4.489(1) & 7.572(1) & 10.364(1) & 6.332(1) & 4.796(1)  & 5.048(1) & 3.931(1)
\\
{$\delta^{\text{NLO}}, \%$}
                     & 9.63(1)  & 19.19(1) & 3.38(1)   & 4.03(1)  & 17.92(1)  & 4.53(1) & 16.98(1)
\\
{$\delta^{\rm QED}, \%$}
                     & 7.63(1)  & 7.52(1)  & 7.50(1)   & 7.57(1)  & 7.60(1)  & 7.61(1)  & 7.65(1)
\\
{$\delta^{\rm weak}, \%$}
                     & 2.01(1)  & 11.68(1) &$-4.12(1)$&$-3.55(1)$ & 10.31(1) & $-3.08(1)$& 9.32(1)
\\
\hline
\multicolumn{8}{|c|}{$\sqrt{s} = 500$~GeV}\\
\hline
$\sigma^{\text{Born}}$, pb
                     & $0.8335(1)$  &$1.2932(1)$ & $2.0407(1)$& 1.2391(1)  & 0.82795(1) & 0.98299(1) & 0.68398(1)
\\
{$\sigma^{\text{NLO}}$, pb}
                     & $0.8801(1)$ & $1.5362(1)$ & $1.9811(1)$& 1.2137(1)  & 0.9692(1)  & 0.9696(1) & 0.7917(1)
\\
{$\delta^{\text{NLO}}, \%$}
                     & $5.65(1)$   & $18.79(1)$ & $-2.92(1)$  & $-2.05(1)$ & 17.06(1)   & $-1.37(1)$ & 15.75(1)
\\
{$\delta^{\rm QED}, \%$}
                     & 7.35(1)     & 7.21(1)    &  7.20(1)    & 7.26(1)    & 7.33(1)    & 7.33(1) & 7.40(1)
\\
{$\delta^{\rm weak}, \%$}
                     & $-1.69(1)$  & 11.60(1)   & $-10.12(1)$ & $-9.33(1)$ & 9.72(1)    & $-8.69(1)$ & 8.36(1)
\\
\hline
\multicolumn{8}{|c|}{$\sqrt{s} = 1000$~GeV}\\
\hline
$\sigma^{\text{Born}}$, pb
                     & 0.19860(1)  &0.30813(1) & 0.48625(1)&0.29524(1) & 0.19728(1) & 0.23422(1) & 0.16297(1)
\\
{$\sigma^{\text{NLO}}$, pb}
                     & 0.19747(3)  & 0.3663(1) & 0.4222(1) &0.260154(4)& 0.22927(2)& 0.20861(2) & 0.18625(2)
\\
{$\delta^{\text{NLO}}, \%$}
                     & $-0.57(2)$  & 18.85(1)  & $-13.17(1)$&$-11.88(1)$ &16.22(1)&$-10.93(1)$ & 14.28(1)
\\
{$\delta^{\rm QED}, \%$}
                     & 7.56(1)     & 7.38(1)   & 7.39(1)    & 7.51(1) & 7.52(1) & 7.55(1) & 7.58(1)
\\
{$\delta^{\rm weak}, \%$}
                     & $-8.13(1)$  & 11.47(1)  & $-20.55(1)$& $-19.38(1)$ & 8.71(1) & $-18.45(1)$ & 6.70(1)
\\
\hline
\end{tabular}
\label{eeazgf0B}
\end{table}


Table~2 presents 
the result of calculating the Born and one-loop cross sections
as well as the relative correction
$\delta=\sigma^{\rm one-loop}/\sigma^{\rm Born}-1$ in percent
of the $e^+e^- \to \gamma Z(\gamma)$ scattering
for c.m. energies $\sqrt{s} = $ $250$, $500$ and $1000$ GeV
in the $\alpha(0)$ scheme.

It is seen that the QED RCs are about $+7.5\%$ and almost constant for the c.m. energies (\ref{set_energy}),
while the weak relative corrections strongly depend on the energy and the degree of the initial beam polarization:
they are mostly positive for  positive electron polarization and mostly negative for  negative electron polarization.
The complete EW NLO corrections also strongly depend on the energy and the degree of polarization.

\subsubsection{Scheme dependence}

To assess the theoretical uncertainty, we perform calculations
in the $\alpha(0)$ and $G_\mu$ EW schemes.
The integrated cross sections for the weak corrections in these schemes and their relative difference
\begin{eqnarray}
\delta_{G_\mu/\alpha(0)} = \frac{\sigma_{G_\mu}}{\sigma_{\alpha(0)}} - 1,\, \%
\label{rgmual0}
\end{eqnarray}
are presented in Table~3.

As is well known, the difference between two EW schemes in the LO is just the
ratio of the EW couplings and gives $\delta^{\rm LO}_{G_\mu/\alpha(0)} = 6.5\%$. 
As seen in the table, the weak contribution reduces the difference
to about
0.55\% at the energy of 250 GeV, 0.32\% at 500 GeV and 0.11\% at 1000 GeV. 
These ratios~(\ref{rgmual0}) 
show stabilization of the results and can be considered as
an estimation of the theoretical uncertainty of weak contributions, 
that is, additional corrections of two and more loops.

\begin{table}[ht]
\caption{
Integrated Born and weak contributions 
to the cross section corrections in two EW schemes,
$\alpha(0)$ and $G_\mu$,
at the center-of-mass energies
(\ref{set_energy})
}
\centering
\begin{tabular}{|l|c|c|c|}
\hline
$\sqs$, GeV & 250         & 500       & 1000\\
\hline
{\sba, pb}  & 4.09449(1) & 0.83348(1)  & 0.19860(1) \\
{\sbg, pb}  & 4.36105(1) & 0.88774(1)  & 0.21152(1) \\
{\dbg, \%}  & 6.51(1)    & 6.51(1)     & 6.50(1) \\
{\swa, pb}  & 4.17661(1) & 0.81936(1)  & 0.18245(1) \\
{\swg, pb}  & 4.19941(1) & 0.82199(1)  & 0.18225(1) \\
{\dwg, \%}  & 0.55(1)    & 0.32(1)     & 0.11(1) \\
\hline
\end{tabular}
\end{table}

\subsection{Differential distributions}

\subsubsection{Angular dependence}

The angular dependence of the unpolarized cross sections for the Born and
one-loop levels in the $\alpha(0)$ EW scheme (upper panel)
and QED and weak relative corrections (lower panel)
is illustrated in Figs.~1-3.
The symbol $\vartheta_Z$ denotes the angle between the initial
positron $e^+(p_1)$ and the $Z$-boson.
This angular curve is symmetrical about zero.
For all c.m. energies
the minimum of the Born and one-loop cross
sections is at zero
while the maximum is in the corners of the angular cut value. 

At $\sqs = 250$ GeV the
QED relative corrections dominate and are only modified slightly by weak corrections.
But as the energy increases, the situation changes.
We observe that at $\sqs =$ 500 and 1000 GeV both corrections
are large, and the strong compensation occurs.

\begin{figure}[h!]
\begin{center}
\includegraphics[width=0.9\linewidth]{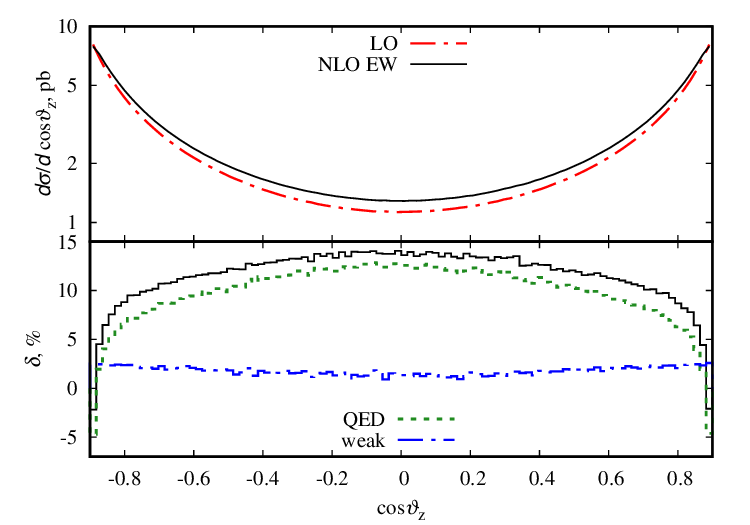}
\caption{
LO and EW NLO (in parts) cross sections and relative corrections
at $\sqs = 250$~GeV
for unpolarized initial beams}
\label{fig:250-cos3}
\end{center}
\end{figure}

\begin{figure}[h!]
\begin{center}
\includegraphics[width=0.9\linewidth]{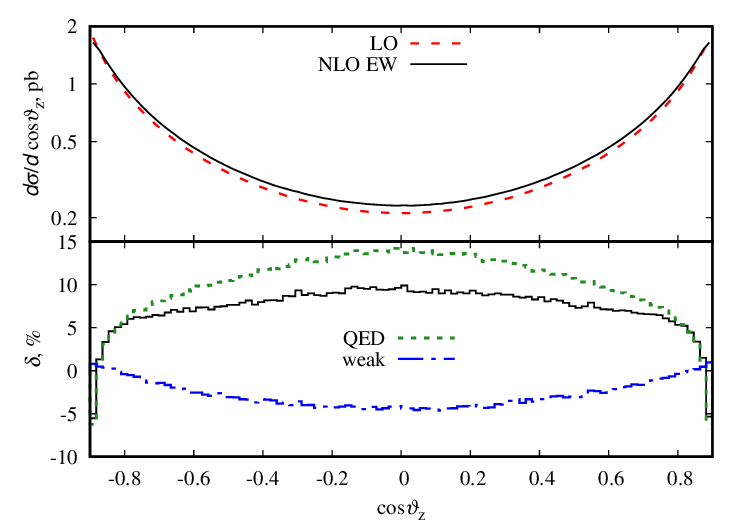}
\caption{
The same as in Fig.~1 but for $\sqs = 500$~GeV}
\label{fig:500-cos3}
\end{center}
\end{figure}

\begin{figure}[!h]
\begin{center}
\includegraphics[width=0.9\linewidth]{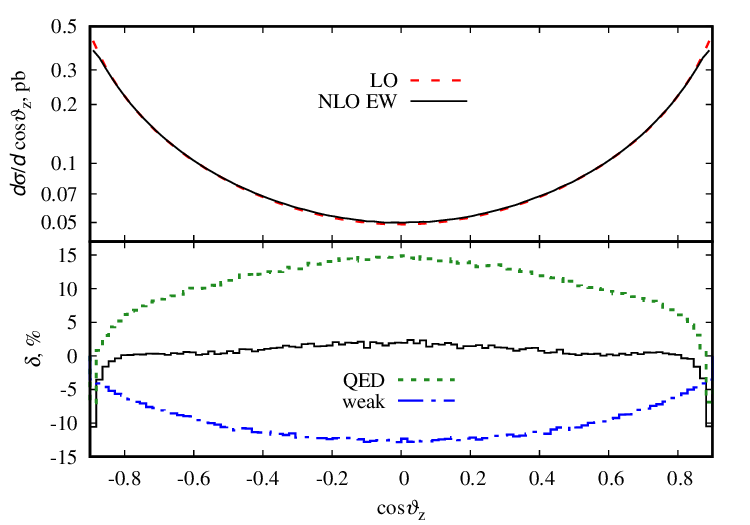}
\caption{
The same as in Fig.~1 but for $\sqs = 1000$~GeV
}
\label{fig:1000-cos3}
\end{center}
\end{figure}

\begin{figure}
\begin{center}
\includegraphics[width=0.9\linewidth]{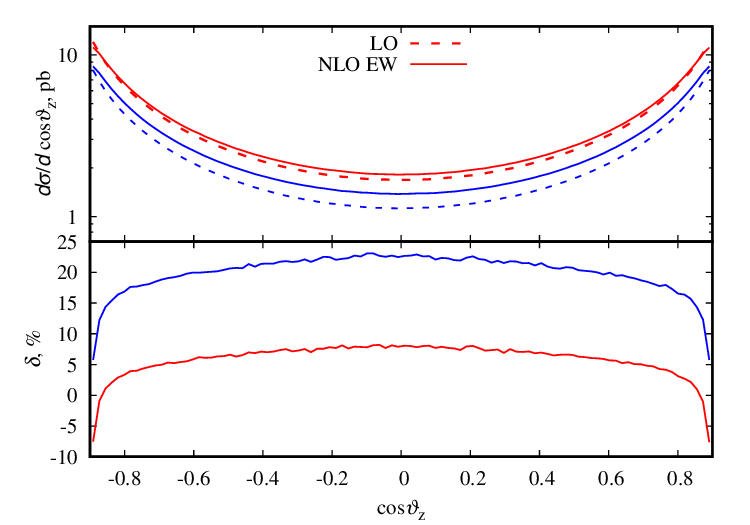}
\caption{
LO and EW NLO cross sections and relative corrections
at $\sqrt{s} = 250$~GeV 
with polarized initial beams.
}
\label{fig_xcross_250_pol}
\end{center}
\end{figure}

\begin{figure}
\begin{center}
\includegraphics[width=0.9\linewidth]{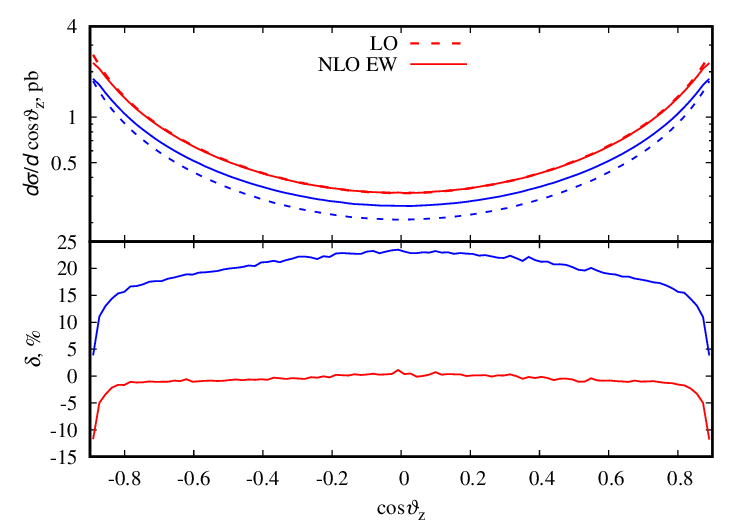}
\caption{
The same as in Fig.~4 but for $\sqs = 500$~GeV
}
\label{fig_xcross_500_pol}
\end{center}
\end{figure}

\begin{figure}
\begin{center}
\includegraphics[width=0.9\linewidth]{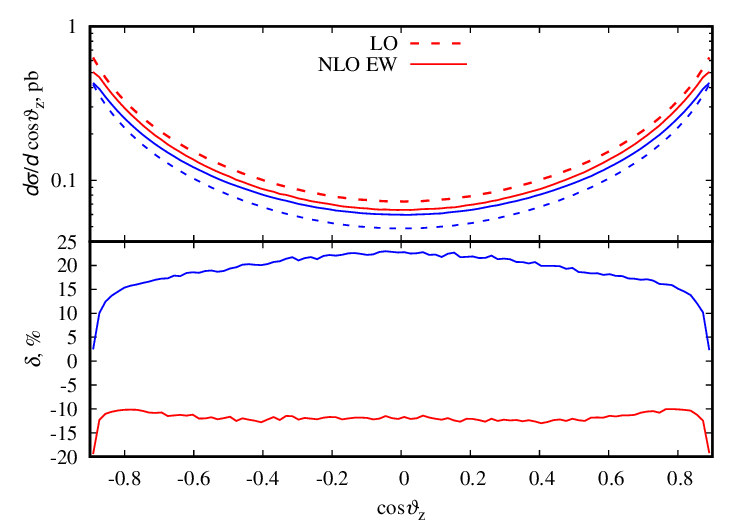}
\caption{
The same as in Fig.~4 but for $\sqs = 1000$~GeV
}
\label{fig_xcross_1000_pol}
\end{center}
\end{figure}

In Figs.~4-6,
the LO (dashed line) and NLO EW (solid line) cross sections
(upper panel) as well as the relative corrections (lower panel)
are shown.
The red lines correspond (solid/dashed lines) to partially polarized initial beams with
$(P_{e^+},P_{e^-}) = (\pm 0.3,\mp 0.8)$.
Radiative corrections significantly reduce cross sections at the energy $\sqrt{s} = $ 250 GeV
in the whole range of scattering angles.
The real planned polarized states in the ILC experiment show a significant
dependence on the polarization of the initial beams, namely,
for
$(P_{e^+},P_{e^-}) = (+0.3,-0.8)$
the relative corrections are from $-10$ up to  $+5\%$ while for $(-0.3,+0.8)$ they are from $5$ to $23\%$.

At the c.m. energy $\sqrt{s} = $ 500 GeV the dependence on polarization is also strong,
and $\delta$ is from $-10$\% to $0$\%.
At the c.m. energy $\sqrt{s} = $ 1000 GeV  $\delta$ varies from $-20$\% to $-10$\% for $(+0.3,-0.8)$.

For $(P_{e^+},P_{e^-}) = (-0.3,+0.8)$ the radiative corrections vary from $5$\% to $23$\% and slightly
depend on the energy.

It should also be noted that the nonphysical dips in the first and last bins of the relative correction
histograms are
due to the angular limits and can be removed by applying wider cuts.

\subsubsection{Energy dependence}
 
The LO and NLO EW corrected unpolarized cross sections in pb
as a function of the c.m. energy are shown in Fig.~7.
At the bottom of the figure the specific contributions of the relative NLO EW corrections are drawn. We decompose the complete one-loop contribution into
two gauge-invariant subsets of the
QED and weak diagrams.
\begin{figure}[h!]
\begin{center}
\includegraphics[width=0.9\linewidth]{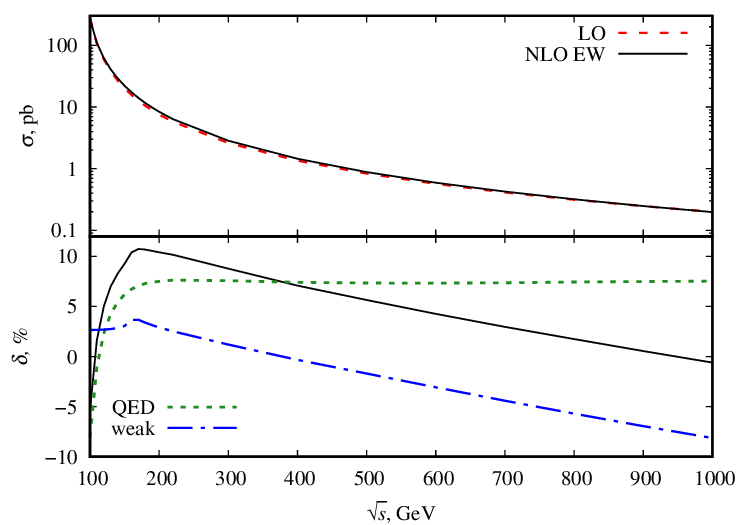}
\end{center}
\caption{
The LO and NLO EW corrected unpolarized cross sections 
and the relative corrections in parts 
as a function of the center-of-mass energy.
}
\label{fig:eeZZdelta}
\end{figure}

In the c.m. energy range from 100 to 200 GeV, the QED corrections have a sharp rising
from negative (about $-10$\%) to positive (about 6\%) values and then are practically constant
while the weak corrections have a pick corresponding to
the under threshold two $W$-boson
box and then start to decrease.

\subsection{Left-right asymmetry}

The left-right asymmetry shows an order of
parity violation and, by definition~(\ref{alr}),
does not depend on the degrees of the initial beam polarization:
\bqa
\label{alr}
\alr = \frac{\sigma_{LR}-\sigma_{RL}}{\sigma_{LR}+\sigma_{RL}},
\eqa
with $\sigma_{LR}$ and $\sigma_{RL}$ being the cross sections for the fully polarized
electron-positron $e^-_Le^+_R$ and $e^-_Re^+_L$ initial states, respectively.
Figure~8 shows the distributions of 
$\alr$ in $\cvz$
for the Born and one-loop contribution at the c.m. energies $\sqrt{s} = 250, 500, 1000$ GeV
in the $\alpha(0)$ EW scheme.
The corresponding shift of the asymmetry
$$\Delta \alr = \alr ({\rm NLO\ EW}) - \alr ({\rm LO})$$
is shown in the lower panel.

At Born level, $\alr$ is constant:
\bqa
\alr^{\rm Born} = \frac{(1-2\sin^2{\theta_W})^2-4\sin^4{\theta_W}}{(1-2\sin^2{\theta_W})^2+4\sin^4{\theta_W}} \approx 0.2243.
\eqa
Here the sine of Weinberg angle is $\sin^2{\theta_W} = 1-{\mw^2}/{\mz^2}$.

\begin{center}
\begin{figure}[!ht]
\centerline{\includegraphics[width=0.9\linewidth]{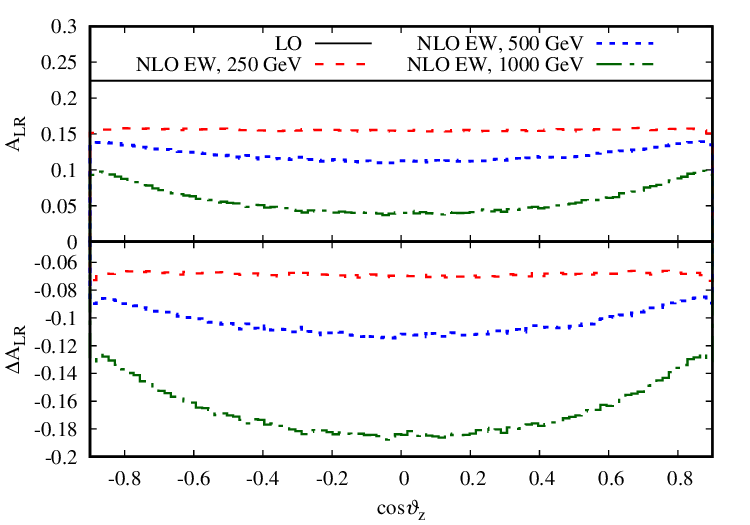}}
\caption{
Distributions of left-right asymmetry $A_{LR}$ in $\cvz$ 
for the Born and one-loop levels for c.m. energies $\sqrt{s}$ = 250, 500, 1000 GeV
in the $\alpha(0)$ EW scheme.
}
\label{alrfigs}
\end{figure}
\end{center} 

The asymmetry at the  c.m. energy $\sqrt{s} = 250$ GeV has a rather flat behavior while at
$500$ and $1000$ GeV there is some increase at $\cvz$ = 0.9 value (the cut value)
compared to the case at $\cvz$ = 0.

\section{Conclusion}\label{sect_Concl}

In this paper, we have described the evaluation of the polarization effects
of the process $e^+e^- \to \gamma Z$ 
at the one-loop level at high energies for future  $e^+e^-$ colliders.

The calculated polarized tree-level cross sections
for the Born and hard photon bremsstrahlung were
compared with the {\calchep} and {\whizard} results and
very good agreement was found.

The effect of the polarization of the initial beams
has been carefully analyzed for some states
both for cross sections and for angular and energy distributions at the one-loop EW level,
and it was found to be significant.
The cross sections were found to depend strongly on the degree of polarization of the original beams.
The radiative corrections themselves were rather sensitive to the degree of polarization of the initial beams
and depended quite strongly on energy.
In addition, calculations in the $\alpha(0)$ and $G_\mu$ EW schemes were considered. 
By combining different experimental criteria, polarization degrees of beams, and the EW schemes, 
radiative corrections can be minimized.

We observe a symmetric
behavior of $\alr$ with respect to $\cos \vartheta_Z = 0$,
the clear signature, the significant dependence on
energy, the flatter behavior with decreasing energy
and the large sensitivity to electroweak interaction effects.

\subsection*{Acknowledgments}
The research was supported by the Russian Foundation for Basic Research, project No. 20-02-00441 (purchasing a computing cluster)
and the Russian Science Foundation, project No. 22-12-00021.
We are grateful to Prof. A.~Arbuzov 
for the help in preparation of the manuscript.


\begingroup\begin{thebibliography}{10}

\bibitem{Blondel:2018aan}
A.~Blondel and P.~Janot,
\href{http://www.arXiv.org/abs/1809.10041}{{\tt 1809.10041}}.

\bibitem{Blondel:2019qlh}
A.~Blondel, A.~Freitas, J.~Gluza, T.~Riemann, S.~Heinemeyer, S.~Jadach, and
  P.~Janot,
\href{http://www.arXiv.org/abs/1901.02648}{{\tt 1901.02648}}.

\bibitem{LEP}
L.~The LEP Collaborations~ALEPH, DELPHI, OPAL, and the LEP TGC Working~Group.

\bibitem{Blondel:2018mad}
A.~Blondel {\em et al.}, ``{Standard Model Theory for the FCC-ee: The
  Tera-Z}'', in {\em {Mini Workshop on Precision EW and QCD Calculations for
  the FCC Studies : Methods and Techniques CERN, Geneva, Switzerland, January
  12-13, 2018}}, 2018,
\href{http://www.arXiv.org/abs/1809.01830}{{\tt 1809.01830}}.

\bibitem{FCC:2018byv}
{FCC} Collaboration, A.~Abada {\em et al.}, {\em Eur. Phys. J. C} {\bf 79}
  (2019), no.~6 474.

\bibitem{ILC:2013jhg}
{ILC} Collaboration, \href{http://www.arXiv.org/abs/1306.6352}{{\tt
  1306.6352}}.

\bibitem{Fujii:2015jha}
K.~Fujii {\em et al.},
\href{http://www.arXiv.org/abs/1506.05992}{{\tt 1506.05992}}.

\bibitem{ILC:2019gyn}
{ILC} Collaboration, H.~Aihara {\em et al.},
  \href{http://www.arXiv.org/abs/1901.09829}{{\tt 1901.09829}}.

\bibitem{CLICdp:2018esa}
{CLICdp} Collaboration, H.~Abramowicz {\em et al.}, {\em JHEP} {\bf 11} (2019)
  003, \href{http://www.arXiv.org/abs/1807.02441}{{\tt 1807.02441}}.

\bibitem{CEPCStudyGroup:2018ghi}
{CEPC Study Group} Collaboration, M.~Dong {\em et al.},
  \href{http://www.arXiv.org/abs/1811.10545}{{\tt 1811.10545}}.

\bibitem{Duan:2023lyp}
Z.~Duan, T.~Chen, J.~Gao, D.~Ji, X.~Li, D.~Wang, J.~Wang, Y.~Wang, and W.~Xia,
  {\em JACoW} {\bf eeFACT2022} (2023) 97--102.

\bibitem{MoortgatPick:2005cw}
G.~Moortgat-Pick {\em et al.}, {\em Phys. Rept.} {\bf 460} (2008) 131--243,
\href{http://www.arXiv.org/abs/hep-ph/0507011}{{\tt hep-ph/0507011}}.

\bibitem{Bambade:2019fyw}
P.~Bambade {\em et al.}, \href{http://www.arXiv.org/abs/1903.01629}{{\tt
  1903.01629}}.

\bibitem{Bondarenko:2022xmc}
S.~Bondarenko, Y.~Dydyshka, L.~Kalinovskaya, A.~Kampf, L.~Rumyantsev,
  R.~Sadykov, and V.~Yermolchyk, {\em Phys. Rev. D} {\bf 107} (2023), no.~7
  073003, \href{http://www.arXiv.org/abs/2211.11467}{{\tt 2211.11467}}.

\bibitem{Mizuno:2021dxw}
T.~Mizuno, K.~Fujii, and J.~Tian, {\em AIP Conf. Proc.} {\bf 2319} (2021),
  no.~1 100004.

\bibitem{Aihara:1995iq}
H.~Aihara {\em et al.}, ``{Anomalous gauge boson interactions}'', in {\em
  {Electroweak symmetry breaking and new physics at the TeV scale}} (T.~L.
  Barklow, S.~Dawson, H.~E. Haber, and J.~L. Siegrist, eds.), 3, 1995,
  \href{http://www.arXiv.org/abs/hep-ph/9503425}{{\tt hep-ph/9503425}}.

\bibitem{Mizuno:2022xuk}
T.~Mizuno, K.~Fujii, and J.~Tian, ``{Measurement of $A_{LR}$ using radiative
  return at ILC 250}'', in {\em {Snowmass 2021}}, 3, 2022,
  \href{http://www.arXiv.org/abs/2203.07944}{{\tt 2203.07944}}.

\bibitem{Bardin:2007dz}
D.~Bardin, S.~Bondarenko, L.~Kalinovskaya, G.~Nanava, L.~Rumyantsev, and W.~von
  Schlippe, {\em Eur. Phys. J. C} {\bf 54} (2008) 187--197, [Erratum:
  Eur.Phys.J.C 82, 417 (2022)], \href{http://www.arXiv.org/abs/0710.3083}{{\tt
  0710.3083}}.

\bibitem{MCSANCee:2021}
A.~B. Arbuzov {\em et al.}, {\em Comput. Phys. Commun. (to be published)}
  (2021).

\bibitem{SADYKOV2020107445}
R.~Sadykov and V.~Yermolchyk, {\em Computer Physics Communications} {\bf 256}
  (2020) 107445, \href{http://www.arXiv.org/abs/2001.10755}{{\tt 2001.10755}}.

\bibitem{CapdequiPeyranere:1984sj}
M.~Capdequi~Peyranere, Y.~Loubatieres, and M.~Talon, {\em Nuovo Cim.} {\bf A90}
  (1985)
363.

\bibitem{Berends:1986yy}
F.~A. Berends, G.~J.~H. Burgers, and W.~L. van Neerven, {\em Phys. Lett.} {\bf
  B177} (1986)
191--194.

\bibitem{Bohm:1986mz}
M.~Bohm and T.~Sack, {\em Z. Phys.} {\bf C35} (1987)
119.

\bibitem{Gounaris:2002za}
G.~J. Gounaris, J.~Layssac, and F.~M. Renard, {\em Phys. Rev.} {\bf D67} (2003)
  013012,
\href{http://www.arXiv.org/abs/hep-ph/0211327}{{\tt hep-ph/0211327}}.

\bibitem{Belyaev:2012qa}
A.~Belyaev, N.~D. Christensen, and A.~Pukhov, {\em Comput. Phys. Commun.} {\bf
  184} (2013) 1729--1769,
\href{http://www.arXiv.org/abs/1207.6082}{{\tt 1207.6082}}.

\bibitem{Kilian:2007gr}
W.~Kilian, T.~Ohl, and J.~Reuter, {\em Eur. Phys. J.} {\bf C71} (2011) 1742,
\href{http://www.arXiv.org/abs/0708.4233}{{\tt 0708.4233}}.

\bibitem{Kilian:2018onl}
W.~Kilian, S.~Brass, T.~Ohl, J.~Reuter, V.~Rothe, P.~Stienemeier, and M.~Utsch,
  ``{New Developments in WHIZARD Version 2.6}'', in {\em {International
  Workshop on Future Linear Collider (LCWS2017) Strasbourg, France, October
  23-27, 2017}}, 2018,
\href{http://www.arXiv.org/abs/1801.08034}{{\tt 1801.08034}}.

\end{thebibliography}\endgroup

\providecommand{\href}[2]{#2}
\end{document}